\documentstyle[12pt]{article}

\begin{document}

\title{Geometric Models of the Relativistic Harmonic Oscillator}

\author{Ion I. Cot\u aescu\\ The West University of Timi\c soara,\\
         V. P\^ arvan Ave. 4, RO-1900 Timi\c soara, Romania}

\date{\today}
\maketitle

\begin{abstract}
A family of relativistic geometric models is defined as a 
generalization of the actual anti-de Sitter (1+1) model of the relativistic
harmonic oscillator. It is shown that all these models lead to the usual 
harmonic oscillator in the non-relativistic limit, even though their 
relativistic behavior is quite different. Among quantum models we find a set 
of models with countable energy spectra, and another one having only a finite 
number of energy levels and in addition a continuous spectrum. 
\end{abstract}

\

The problem of the relativistic generalization of the 
(classical or quantum) non-relativistic harmonic oscillator (NRHO) 
has been frequently discussed, but as yet there is no unique definition 
of the relativistic harmonic oscillator (RHO). In the context of 
general relativity, the RHO has been defined as a free system on the  
anti-de Sitter (AdS) static background. There are many phenomenologic 
\cite{P1,P2} and group theoretic
\cite{A1,A2,A3} arguments for this geometric model. Its advantage is  
that the constants of the classical motion (on geodesics) satisfy  the 
$so(1,2)$ algebra (of the AdS symmetry \cite{A1}), realized by Poison brackets for 
any AdS metric. However, the concrete choice of the metric is also important 
from the observer's point of view because the form of the classical trajectory, 
as well as the modes of the corresponding quantum system, depend on 
it. For the (1+1) RHO the (1+3) metric proposed in Ref. \cite{P2}  gives 
the following (1+1) AdS metric 
\begin{equation}\label{(m1)}
ds^{2}=\frac{1}{1-\omega^{2} x^{2}}dt^{2}-\frac{1}{(1-\omega^{2}
x^{2})^{2}}dx^{2},
\end{equation}
which reproduces the classical non-relativistic equation of motion, i.e. 
$\ddot x +\omega^{2}x=0$. The corresponding quantum system has been analyzed 
in Ref. \cite{M}, starting with another AdS metric which can be derived from 
(\ref{(m1)}) by  changing the space coordinate  
$x \rightarrow x'=x/\sqrt{1+\omega^{2}x^{2}}$.
The result is an equidistant energy spectrum with a groundstate energy larger 
than, but approaching $\omega/2$ in the non-relativistic limit (in natural 
units, $\hbar =c =1$). Since, partularlly, the space coordinate transformations 
of the static backgrounds do not change the quantum modes, we can say that 
the RHO is well simulated by the free motion on the AdS background with the 
metric (\ref{(m1)}).

However, one can ask if there are  more geometric models which should 
behave as the NRHO in the non-relativistic limit. In order to give an answer, 
we  shall study here the  family of models having the metrics 
\begin{equation}\label{(m)}
ds^{2}=g_{00}dt^{2}+g_{11}dx^{2}=
\frac{1+(1+\lambda) \omega^{2}x^{2}}{1+\lambda \omega^{2} x^{2}}dt^{2}-  
\frac{1+(1+\lambda) \omega^{2}x^{2}}{(1+\lambda \omega^{2} x^{2})^{2}}dx^{2},
\end{equation}
where $\lambda$ is a real parameter. Our aim is to investigate both the 
classical and the quantum free motions of a particle of mass $m$,   
and to show that all these  models lead to the NRHO in the non-relativistic 
limit. Moreover, we shall try to point out their specific relativistic effects. 
 
The metrics (\ref{(m)}) represent a generalization of the AdS metric 
(\ref{(m1)}). In fact these are  conformal transformations depending on 
$\lambda$ of some 
AdS metrics if $\lambda<0$, of some static de Sitter metrics 
if  $\lambda>0$, or of the the Minkowski flat metric when $\lambda=0$. 
The parameterization of these transformations has been defined 
in a such a manner to obtain the exact AdS metric (\ref{(m1)})  
for $\lambda = -1$. We note that the event horizon of an observer situated at 
$x=0$ is at $R_{+}=\infty$ for $\lambda\geq 0$ and at $R_{-}=1/\omega \sqrt{-\lambda}$ in 
the case of $\lambda<0$ in which the metrics have singularities. This will give 
the space domain of the free motion, $D=(-R,R)$.

First, we shall derive the classical equation of motion starting with the 
geodesics equation, (non-covariantly) expressed in terms of $x(t)$ 
and its time derivatives $\dot x$ and $\ddot x$,
\begin{equation}\label{(geo)}
\ddot x -\frac{g_{00,x}}{2g_{11}}+ (\dot x)^{2}\left( \frac{g_{11,x}}{2g_{11}}-
\frac{g_{00,x}}{g_{00}} \right)=0.
\end{equation}
In addition we shall use the conservation of the energy $E$ (on the static 
backgrounds), which gives
\begin{equation}\label{(con)}
(\dot x)^{2}=\frac{g_{00}}{g_{11}} \left( \frac{m^{2}g_{00}}{E^{2}}-1 \right).
\end{equation}
From (\ref{(m)}), (\ref{(geo)}) and (\ref{(con)}) we obtain
\begin{equation}\label{(mot)}
\ddot x + \Omega^{2} x=0,
\end{equation}
where
\begin{equation}\label{(freq)}
\Omega=\frac{\omega}{E} \sqrt{(1+\lambda)m^{2}-\lambda E^{2}}
\end{equation}
is the effective frequency. Its dependence on energy can be considered as a 
pure relativistic effect. We note that $\Omega$ does not depend on $E$ only    
in the case of the AdS metric when $\lambda=-1$. The trajectory,   
\begin{equation}\label{(x)}
x(t)=a\sin(\Omega (t-t_{0})),
\end{equation}
is one of oscillations if $\Omega$ and the amplitude,
\begin{equation}
a=\frac{1}{\omega}\left(\frac{E^{2}-m^{2}}{(1+\lambda)m^{2}-\lambda E^{2}}
\right)^{\frac{1}{2}}
\end{equation}
are real numbers. We observe  that for $\lambda>0$ and $E^{2}\geq 
m^{2}(1+1/\lambda)$ 
these oscillations degenerate into open (uniform or accelerated) motions on 
$D=(-\infty, \infty)$.   
Hence, for $\lambda>0$ the system oscillates only for  
$E\in [m, m\sqrt{1+1/\lambda})$. This could lead to a finite discrete energy 
spectrum for the quantum motion. However, when  
$\lambda<0$ the system oscillates for all the possible energies, $E\in [m, 
\infty)$, with  amplitudes remaining less than $R_{-}$. Now, we can verify that 
in the non-relativistic limit, 
for very small $E_{nr}=E-m$ and for any $\lambda$, we obtain the familiar 
expressions $\Omega \rightarrow \omega$ and $a^{2}\rightarrow 
2E_{nr}/m^{2}\omega^{2}$. 
Therefore, at least in the case of the classical motion, all 
these models have as the non-relativistic limit the classical NRHO.  
It remains to verify if this property remains valid also for the quantum 
motion.

The quantum free motion of a spinless particle is described by the scalar 
field  $\phi$ defined on $D$ minimally coupled with the gravitational field 
\cite{B1} given by the metric (\ref{(m)}).
Because of  energy conservation, the Klein-Gordon equation
\begin{equation}\label{(kg)}
\frac{1}{\sqrt{-g}}\partial_{\mu}\left(\sqrt{-g}g^{\mu\nu}\partial_{\nu}\phi
\right) + m^{2}\phi=0
\end{equation}
where $g=\det(g_{\mu\nu})$, admits a set of fundamental solutions (of positive 
and negative frequency) of the form
\begin{equation}\label{(sol)}
\phi_{E}^{(+)}=\frac{1}{\sqrt{2E}}e^{-iEt}U_{E}(x), \quad 
\phi^{(-)}=(\phi^{(+)})^{*},
\end{equation}
which must be orthogonal with respect to the relativistic scalar product 
\cite{B1} 
\begin{equation}\label{(sp1)}
<\phi,\phi'>=i\int_{D}dx\sqrt{-g}g^{00}\phi^{*}\stackrel{\leftrightarrow}{\partial_{0}} \phi'.
\end{equation}
Starting with the metric (\ref{(m)}) we obtain 
the Klein-Gordon equation  
\begin{equation}\label{(kg1)}
(1+\lambda\omega^{2}x^{2})U_{,xx}+\lambda\omega^{2}xU_{,x}+  {E}^{2}U  
-\frac{(1+(1+\lambda)\omega^{2}x^{2})}
{(1+\lambda\omega^{2}x^{2})}m^{2}U=0  
\end{equation}
and the concrete form of the scalar product
\begin{equation}\label{(psc)}
<U,U'>=\int_{D}\frac{dx}{\sqrt{1+\lambda\omega^{2}x^{2}}}U^{*}U'.
\end{equation}
In the following we shall try to derive the energy spectrum and the form of 
the wave functions up to normalization factors.

When $\lambda=0$ the equation (\ref{(kg1)}) becomes
\begin{equation}
-U_{n,xx}+m^{2}\omega^{2}x^{2}U_{n}=({E_{n}}^{2}-m^{2})U_{n},
\end{equation}
from which it results that the wave functions $U_{n}$ coincide with those 
of the NRHO, while the energy spectrum 
\begin{equation}\label{(s1)}
{E_{n}}^{2}=m^{2}+2m\omega(n+\frac{1}{2}), \quad n=0,1,2,...
\end{equation} 
goes to the traditional one in the non-relativistic limit.

In the general case of any $\lambda\not=0$ it is convenient to use the new 
variable $y=-\lambda\omega^{2}x^{2}$, and the notations
\begin{equation}\label{(nu)}
\epsilon= \frac{E}{\lambda\omega}, \quad \mu=\frac{m}{\lambda\omega}, \quad 
\nu=\frac{1}{4}[(1+\lambda)\mu^{2}-\lambda \epsilon^{2}].
\end{equation}
We shall look for a solution of the form
\begin{equation}\label{(100)}
U(y)=N(1-y)^{p}y^{s}F(y),
\end{equation}
where $p$ and $s$ are real numbers and $N$ is the normalization factor. After 
a few manipulation we find that, for  
\begin{equation}\label{(par)}
s(2s-1)=0, \quad 4p^{2}-2p-\mu^{2}=0,
\end{equation}
the equation (\ref{(kg1)}) transforms into the following hypergeometric 
equation:
\begin{equation}
y(1-y)F_{,yy}+[2s+\frac{1}{2}-y(2p+2s+1)]F_{,y}-[(p+s)^{2}-\nu]F=0.
\end{equation}
This has the solution \cite{B2}
\begin{equation}\label{(F1)}
F=F(p+s-\sqrt{\nu}, p+s+\sqrt{\nu}, 2s+\frac{1}{2}, y),
\end{equation}
which depends on the possible values of the parameters $p$ and $s$. From 
(\ref{(par)}) it follows that  
\begin{equation}\label{(p)}
s=0,\frac{1}{2}, \quad p=p_{\pm}=\frac{1}{4}[1\pm \sqrt{1+ 4\mu^{2}}].
\end{equation}
Moreover, when 
\begin{equation}\label{(quant1)}
\nu=(p+s+n')^{2}, \quad n'=0,1,2...,
\end{equation}
$F$ reduces to a polynomial of degree $n'$ in $y$.  
By using these results, we can establish the general form of the 
solutions of (\ref{(kg1)}), which could be square integrable with respect to 
(\ref{(psc)}), namely 
\begin{equation}\label{(U1)}
U_{n',s}(x)=N_{n',s}(1+\lambda\omega^{2}x^{2})^{p}x^{2s}F(-n',2p+2s+n',
2s+\frac{1}{2}, -\lambda\omega^{2}x^{2}).
\end{equation}
Furthermore, we shall define the quantum number $n=2(n'+s)$ which has odd 
values for $s=0$ and even values for $s=1/2$. For both 
sequences, (\ref{(quant1)}), (\ref{(par)}) and (\ref{(nu)}) give the same 
formula of the energy levels,
\begin{equation}\label{(el)}
{E_{n}}^{2}=m^{2}-\lambda\omega^{2}[4p(n+\frac{1}{2})+n^{2}], \quad 
n=0,1,2... .
\end{equation}  
Now, it remains  to fix the suitable values of 
$p$ for which $<U_{n',s},U_{n',s}> < \infty$, and to analyze the structure of 
the obtained spectra. 

Let us first take  $\lambda>0$. In this case $D=(-\infty,\infty)$, and the 
solution (\ref{(U1)}) will be  square integrable only if $p=p_{-}$ and 
$n<-2p_{-}$. This means that the discrete spectrum is {\it finite}, 
with $n=0,1,2...n_{max}$, 
where $n_{max}$ is the integer part of  $(\sqrt{1+4\mu^{2}} - 1)/2$. 
One can verify that this discrete spectrum is included in the domain of 
energies $[m, m\sqrt{1+1/\lambda})$, for which the classical motion is 
oscillatory. On the other hand, when $E>m\sqrt{1+1/\lambda}$, then $\nu$ is 
negative and the hypergeometric functions (\ref{(F1)}) cannot be reduced 
to polynomials, but remain analytic for negative arguments. Therefore the 
functions 
\begin{equation}
U_{\nu,s}=N_{\nu,s}(1+\lambda\omega^{2}x^{2})^{p_{-}}x^{2s}F(p_{-}+
s-\sqrt{\nu},
p_{-}+s+\sqrt{\nu},2s+\frac{1}{2},-\lambda\omega^{2}x^{2})
\end{equation} 
can be interpreted as the non-square integrable solutions corresponding to 
the {\it continuous} energy spectrum $[m\sqrt{1+1/\lambda}, \infty)$.

In the case of $\lambda<0$ the domain $D=(-R_{-},R_{-})$ is finite (as in 
Ref. \cite{P2}) and, therefore, the  polynomial solutions (\ref{(U1)}) will be  
square integrable over $D$ only if $p=p_{+}$. We observe that there are no 
restrictions on the range of $n$ and, consequently, the discrete spectrum will 
be countable. Moreover, in 
this case we have no continuous spectrum since the hypergeometric functions 
(\ref{(F1)}) generally diverge for $y\rightarrow 1$ (when $x\rightarrow R_{-}$).    
We note that for the AdS metric ($\lambda=-1$) we obtain the same result as in 
Ref. \cite{M},  namely $E_{n}=\omega(2p_{+}+n)$.

Thus we have solved the quantum problem for all the values of the parameter 
$\lambda$. Herein it is interesting to observe that our results are continuous 
in $\lambda$. More precisely, in the limit of $\lambda \rightarrow 0$ the 
general formulae  (\ref{(el)}) and (\ref{(U1)}) will give  the energy spectrum 
(\ref{(s1)}) and the NRHO wave functions. Indeed, we observe that, in this 
limit, $p\sim -m/2\lambda \omega$ (since we have choose  $p=p_{-}<0$ for $\lambda>0$ 
and $p=p_{+}>0$ for $\lambda<0$) and $n_{max}\sim m/\lambda\omega\rightarrow 
\infty$. Therefore, the finite discrete spectra of the models with $\lambda>0$  
become countable  while the continuous spectra 
disappear. Hence, it is obvious that all the discrete spectra given by 
(\ref{(el)}) go to (\ref{(s1)}) when $\lambda \rightarrow 0$. Furthermore, 
we can calculate, up to factors, the  limit of the wave 
functions (\ref{(U1)}). We obtain 
\begin{equation}\label{(U2)}
U_{n',s} \rightarrow \sim e^{-m\omega x^{2}/2}x^{2s}F(-n',2s+\frac{1}{2}, 
m\omega x^{2})\sim e^{-m\omega x^{2}/2}H_{n}(\sqrt{m\omega}x),
\end{equation}
where $H_{n}$ are  Hermite polynomials and $n=2(n'+s)$ as  defined above. 
Thus it results that all the functions (\ref{(U1)}) go to the specific 
NRHO wave function which are just the wave functions for $\lambda=0$. For this 
reason the hypergeometric functions of  (\ref{(U1)}) with the factors $x^{2s}$ 
could be considered as a generalization of the Hermite polynomials.

The non-relativistic limit of our models, defined by $m/\omega \rightarrow \infty$, 
can be easily calculated starting with the observation that, according to 
(\ref{(nu)}) and (\ref{(p)}), this is  
equivalent with the limit $\lambda \rightarrow 0$ and, in addition, 
$m \gg \omega$. Hence, the non-relativistic limit of a model with any $\lambda$ 
will be the same as in the case of $\lambda=0$, i.e. the NRHO. Therefore,  
we can conclude that all the models of RHO we have studied here have 
the same non-relativistic limit, even though they are in fact very different. 

Finally we must specify that among these models 
only one has the whole properties of the NRHO. This is that of the AdS metric 
for which: i. the classical motion is oscillatory with a fixed frequency 
(independent on $E$), and, ii. the quantum energy spectrum is equidistant. 
Another interesting model is that of $\lambda=0$ because its wave functions 
coincide with those of NRHO. However, in general, the models with 
$\lambda \not=-1$ can not be considered as  pure harmonic oscillators since 
these have not the above mentioned properties of the AdS model. 
On the other hand, the new models we have considered are interesting 
because their specific countable or finite discrete energy spectra  
could allow to identify new observable relativistic effects.

\end{document}